%% file: main.tex
\documentclass{article}
\usepackage{spconf,amsmath,amsfonts,graphicx,hyperref,mathtools,xcolor,twemojis, bm, bbm}

\def\BibTeX{{\rm B\kern-.05em{\sc i\kern-.025em b}\kern-.08em
    T\kern-.1667em\lower.7ex\hbox{E}\kern-.125emX}}

\mathtoolsset{showonlyrefs=true} 

\title{Learning False Discovery Rate Control via \\ Model-Based Neural Networks}
\name{Arnau Vilella$^{\star}$ \qquad Jasin Machkour$^{\dagger}$ \qquad Michael Muma$^{\dagger}$ \qquad Daniel P. Palomar$^{\star}$
\thanks{A. Vilella (avp@connect.ust.hk) is supported by the Hong Kong PFS 23-90463 fellowship. J. Machkour (jasin.machkour@tu-darmstadt.de) and M. Muma (michael.muma@tu-darmstadt.de) are supported by the ERC Starting Grant ScReeningData (Project Number: 101042407). D.P. Palomar (palomar@ust.hk) is supported by the Hong Kong GRF 16206123 research grant.
}}
  
\address{$^{\star}$ The Hong Kong University of Science and Technology, Hong Kong SAR, China \\
$^{\dagger}$ Technische Universität Darmstadt, Robust Data Science Group, Germany}

\input{packages.tex}
\input{macros.tex}

\begin{document}
\setstretch{0.874} 
\maketitle
\begingroup
\renewcommand\thefootnote{}\footnotetext{
© 2026 IEEE. Personal use of this material is permitted. Permission from IEEE must be obtained for all other uses, in any current or future media, including reprinting/republishing this material for advertising or promotional purposes, creating new collective works, for resale or redistribution to servers or lists, or reuse of any copyrighted component of this work in other works.
 
This work has received funding from the European Research Council (ERC) under the European Union’s Horizon Europe research and innovation programme (Grant agreement No. 101042407 – ScReeningData).
 
For the purpose of Open Access, the author has applied a Creative Commons Attribution (CC BY) licence to any Author Accepted Manuscript version arising from this submission.
}
\addtocounter{footnote}{-1}
\endgroup
\begin{abstract}
Controlling the false discovery rate (FDR) in high-dimensional variable selection requires balancing rigorous error control with statistical power. Existing methods with provable guarantees are often overly conservative, creating a persistent gap between the realized false discovery proportion (FDP) and the target FDR level. We introduce a learning-augmented enhancement of the T-Rex Selector framework that narrows this gap. Our approach replaces the analytical FDP estimator with a neural network trained solely on diverse synthetic datasets, enabling a substantially tighter and more accurate approximation of the FDP. This refinement allows the procedure to operate much closer to the desired FDR level, thereby increasing discovery power while maintaining effective approximate control. Through extensive simulations and a challenging synthetic genome-wide association study (GWAS), we demonstrate that our method achieves superior detection of true variables compared to existing approaches.

\end{abstract}
\begin{keywords}
Variable selection, false discovery rate (FDR) control, model-based neural networks, T-Rex Selector, high-dimensional data
\end{keywords}
\section{Introduction}
\label{sec:intro}

Controlling the false discovery rate (FDR) in high-dimensional variable selection is a fundamental problem in fields like genomics and finance, where the goal is to identify truly influential variables from a vast set of candidates while maximizing the true positive rate (TPR)~\cite{candes2018panning, machkour2025trex}.

The FDR and TPR are defined as follows: Let $\mathcal{A}$ denote the true set of active variables and $\widehat{\mathcal{A}}$  be the set that has been selected by a variable selection method. Then, the FDR and the TPR are defined, respectively, as the expected values of the sample-level false discovery proportion (FDP) and true positive proportion (TPP), i.e.:
\begin{equation}
\label{eq:metrics}
\text{FDP} \coloneqq \frac{|\widehat{\mathcal{A}} \setminus \mathcal{A}|}{\max\{1, |\widehat{\mathcal{A}}|\}}, \quad \text{TPP} \coloneqq \frac{|\mathcal{A} \cap \widehat{\mathcal{A}}|}{\max\{1, |\mathcal{A}|\}}.
\end{equation}

Although methods like the Lasso~\cite{tibshirani1996regression} and elastic net~\cite{zou2005regularization} induce sparsity in high-dimensional settings, they lack explicit FDR control. There exist well-known FDR-controlling methods for the low-dimensional setting, such as the Benjamini-Hochberg method~\cite{benjamini1995control} and the Benjamini-Yekutieli method~\cite{benjamini2001control}. However, these methods are not applicable in the considered high-dimensional setting, where the number of variables $p$ exceeds the number of samples $n$~\cite{machkour2025trex}. In response, rigorous FDR-controlling methods like the knockoff filter~\cite{barber2015controlling, candes2018panning} were developed, but their computational cost can be prohibitive for large-scale problems. The Terminating-Random Experiments (T-Rex) selector~\cite{machkour2025trex, machkour2025datrex, TRexSelectorpackage} was introduced as a scalable and efficient alternative that offers provable FDR control through a fast, dummy-variable-based experimental framework.

The statistical FDR-control guarantee of the T-Rex selector, however, relies on a provably conservative analytical estimator for the FDP. While this ensures rigorous error control, it often leads to an unnecessary loss of statistical power, particularly in low signal-to-noise ratio (SNR) settings. The method can be overly cautious, achieving an FDR far below the target level at the cost of missing true discoveries. This difficulty in achieving high power in low-SNR regimes is a common challenge, as competing FDR-controlling methods often exhibit poor performance under these conditions.

To address this trade-off, we propose a novel approach that replaces the conservative analytical estimator with a more accurate, data-driven one learned by a model-based neural network~\cite{shlezinger2023model, shultzman2023generalization}. By learning a tighter FDP estimator directly from data, our method can calibrate its selection threshold to operate closer to the desired FDR boundary. This data-driven calibration approach \cite{chen2022learning} trades the formal theoretical guarantee for a significant empirical increase in power (TPR), enabling the discovery of more true active variables while maintaining approximate FDR control, typically within a few percentage points of the target, which is a practical trade-off in all but the most stringent applications.

The main contributions of this paper are: (i) We propose a novel, learning-based framework that integrates a deep neural network into the T-Rex Selector's calibration stage to provide a more accurate, data-driven FDP estimate. (ii) We demonstrate through extensive experiments that our method, trained solely on diverse synthetic data, significantly boosts statistical power (i.e., TPR) compared to the original T-Rex Selector while maintaining approximate FDR control. (iii) We demonstrate that the trained model generalizes well to realistic genomics data.

\section{The Terminating-Random Experiments (T-Rex) Selector}
\label{sec:trexselector}
The T-Rex Selector \cite{machkour2025trex, scheidt2023solving,TRexSelectorpackage} is a fast and scalable framework for high-dimensional variable selection that provides provable finite-sample FDR guarantees while maximizing the number of selected variables. The methodology is illustrated in Fig.~\ref{fig:trex_scheme} and briefly revisited in the following.

\input{TRex_framework.tex}

Variable selection in a standard sparse linear regression model, i.e.,
\begin{equation}
\label{eq:linear_model}
\y = \X \bm{\beta} + \bm{\varepsilon},
\end{equation}
requires determining the support of $\bm{\beta}$. To do this, the T-Rex selector runs $K$ parallel random experiments. In each experiment, the original $n \times p$ predictor matrix $\X$ is augmented with $L$ randomly generated dummy variables, creating an enlarged matrix $\XWK_{k} = [ \X~\D_{k} ]$. Valid dummies can be easily generated by sampling i.i.d. random variables from a standard Gaussian distribution \cite{machkour2025trex}. A forward selection algorithm, such as LARS, is applied to the enlarged data $(\XWK_{k}, \y)$ but is terminated early, once a predefined number of dummy variables $T$ have entered the model \cite{TLARSpackage}. The results are then aggregated by calculating the relative frequency $\Phi_{T,L}(j)$ for each original variable, defined as the fraction of experiments in which it was selected, i.e.,
\begin{equation}
\label{eq:relative_frequency}
\Phi_{T,L}(j) \coloneqq \frac{1}{K} \sum_{k=1}^{K} \mathbbm{1}_{\{j \in \mathcal{C}_{k, L}(T)\}},
\end{equation}
where $\mathbbm{1}_{\{ \cdot \}}$ is the indicator function and $\mathcal{C}_{k,L}(T)$ denotes the set of original variables selected in the $k$-th experiment. Variables are included in the final selected active set, $\widehat{\mathcal{A}}$, if their relative occurrence exceeds a voting threshold $v \in [0.5, 1)$. The T-Rex Selector achieves provable FDR control through a rigorously constructed analytical estimator, $\widehat{\mathrm{FDP}}(v, T, L)$. This estimator relies on the calibration parameters $(v, T, L)$ and the deflated relative occurrence $\Phi'_{T,L}(j)$, a transformation that penalizes the original frequencies $\Phi_{T,L}(j)$ based on the proportion of dummies entering the model at each step, correcting for the random interspersion of null variables among active ones and producing a provably conservative estimate:

\begin{equation}
\label{eq:fdp_estimator}
\widehat{\text{FDP}}(v, T, L) \coloneqq \frac{\sum_{j \in \widehat{\mathcal{A}}(v)} (1 - \Phi'_{T, L}(j))}{\max\{1, |\widehat{\mathcal{A}}(v)|\}}.
\end{equation}
It is proven in \cite{machkour2025trex} that $\mathbb{E}[\widehat{\text{FDP}}] \ge \text{FDR}$, which guarantees that enforcing the condition $\widehat{\text{FDP}} \le \alpha$ for a target level $\alpha$ ensures that $\text{FDR} \le \alpha$. This is used by a calibration algorithm that finds the optimal parameters by solving a grid search
\begin{equation}
\label{eq:optimization}
\max_{v, T, L} |\widehat{\mathcal{A}}_{L}(v, T)| \quad \text{s.t.} \quad \widehat{\text{FDP}}(v, T, L) \leq \alpha.
\end{equation}

While this conservativeness provides a strong theoretical guarantee, the gap between the FDP estimator and the true FDP can be substantial, forcing the method to be overly cautious. In fact, the achieved FDR can be far below the target level $\alpha$, leading to a significant loss in TPR as potentially true variables are discarded to satisfy the conservative constraint from \eqref{eq:optimization}.

\section{Proposed: A Learned FDP Estimator}
\label{sec:proposed}

To boost the statistical power (i.e., TPR) of the T-Rex Selector, we replace its conservative analytical FDP estimator from Eq.~\eqref{eq:fdp_estimator} by a neural network that learns a tighter, more accurate FDP approximation. This allows the calibration algorithm to operate closer to the target FDR ($\alpha$) boundary, trading the formal theoretical guarantee for a substantial gain in empirical power while maintaining approximate FDR control. As shown in Fig.~\ref{fig:proposed_scheme}, this learned estimator drives the ``Calibrate \& Fuse'' module.

\input{TRex_framework_enhanced.tex}

The neural network's input is the flattened vector of T-Rex statistics $(\Phi, v, T, L)$. To accommodate varying numbers of predictors $p$, the relative frequency matrix $\Phi$ is zero-padded to a fixed maximum dimension prior to vectorization. This design is consistent with the definition of relative frequency: predictors absent from a problem naturally have frequency zero. In terms of the architecture, we utilize a multi-layer perceptron (MLP) with three ReLU-activated hidden layers (128, 64, and 32 neurons). A final sigmoid activation on the output neuron constrains the predicted FDP to the interval $[0,1]$. The training is performed once offline, and the online inference cost of this lightweight network is negligible compared to generating the relative frequencies $\Phi$, a step that is already required by the original T-Rex Selector.

Training the network requires access to the ground-truth active set $\mathcal{A}$. Real-world data are unsuitable for two reasons. First, labeled datasets with known $\mathcal{A}$ are rarely available in exploratory applications where variable selection is most critical. Second, training directly on large real datasets would confer an unfair advantage: the network would compress problem-specific information, effectively inflating the sample size and reducing the challenge of the high-dimensional regime ($p \gg n$). To ensure fairness and generalizability, we train exclusively on synthetically generated data. This synthetic setting provides $\mathcal{A}$ by construction, yielding exact FDP labels while forcing the network to learn a broadly applicable calibration function rather than memorizing domain-specific structure.

Our data generation pipeline constructs diverse synthetic systems as follows. A feature matrix $\X$ is drawn from a variety of distributions, and a sparse coefficient vector $\boldsymbol{\beta}$ defines the true active set $\mathcal{A}$. The response vector $\y$ is then generated from the linear model in Eq.~\eqref{eq:linear_model}, with varying signal-to-noise ratios. For each realization, we run $K$ T-Rex experiments to compute the relative frequency matrix $\Phi$. For a fixed number of dummies $L$, a single training instance is formed by sampling calibration parameters $(v, T)$ and computing the true FDP from $\mathcal{A}$. This value serves as the label for the corresponding input $(\Phi, v, T, L)$.

After training, the learned estimator $\widehat{\text{FDP}}$ replaces the analytical estimator in the T-Rex Selector. Calibration is performed by grid search, as in \eqref{eq:optimization}, to identify the optimal $(v^\star, T^\star)$, while $L$ is retained from the original T-Rex Selector. The final active set then becomes
$\widehat{\mathcal{A}}_{L}(v^{\star}, T^{\star})= \{ j : \Phi_{T^\star, L}(j) > v^\star \}$.

The original T-Rex Selector achieves control by systematically overestimating the FDP in expectation. To encourage caution without becoming too conservative, we introduce an asymmetric loss that penalizes underestimation more heavily:
\begin{equation}
\label{eq:asymmetric_loss}
\mathcal{L}(\hat{\xi}, \xi) = \max(0, \hat{\xi} - \xi)^2 + w \cdot \max(0, \xi - \hat{\xi})^2, \quad w > 1.
\end{equation}
Although this does not yield a formal guarantee, it empirically promotes conservative FDP estimates and aligns with prior evidence on asymmetric scoring rules \cite{patton2007properties}.

\section{Artificial data experiments}
\label{sec:art_exp}

To evaluate our method, we use exclusively synthetic data, which provides the known ground-truth active set $\mathcal{A}$ required for precise FDP and TPR calculations. Our primary goal is to assess the model's ability to generalize. To this end, the network is trained on a large, diverse dataset of 1.4 million synthetic systems where each feature matrix $\mathbf{X}$ is drawn from one of fourteen distributions (Beta, Binomial, Cauchy, Chi-Squared, Exponential, Gamma, Gaussian, Gumbel, Laplace, Log-Normal, Pareto, Student's t, Uniform, and Weibull), with randomized hyperparameters to ensure variety. This diverse training promotes generalization to unseen distributions by preventing the model from memorizing statistical properties specific to any single one. This also ensures a fair comparison with the T-Rex Selector, as the training process is agnostic to the test distribution, preventing the network from gaining an unfair advantage that would be analogous to artificially increasing the problem's sample size. For each synthetic system, training entries $(\Phi, v, T, L, \text{FDP})$ are extracted as described in Section~\ref{sec:proposed}.

Initial feasibility studies confirmed the approach's viability: a network trained only on Gaussian data performed well when tested on both seen (Gaussian) and unseen (Student's t) distributions. These studies also served to set the asymmetric loss weight to $w=1.1$. For a more rigorous generalization test, the model was later trained on all fourteen distributions and evaluated on a held-out test set of 10,000 systems from Gaussian Mixture Models (GMMs), chosen for their complexity and dissimilarity to the training data.

The main experiment used synthetic systems with $p=150$ predictors, $n=75$ samples, $|\mathcal{A}|=3$ active variables, and a signal-to-noise ratio (SNR) ranging from $0.01$ to $5$. We set T-Rex parameters to $K=20$ experiments. The network was trained for 10 epochs with a learning rate of $10^{-3}$. Full implementation details are available online.\footnote{\url{https://github.com/ArnauVilella/learning-fdr-control}}

\begin{figure*}[htb]
  \centering
  \centerline{\includegraphics[width=0.7\linewidth]{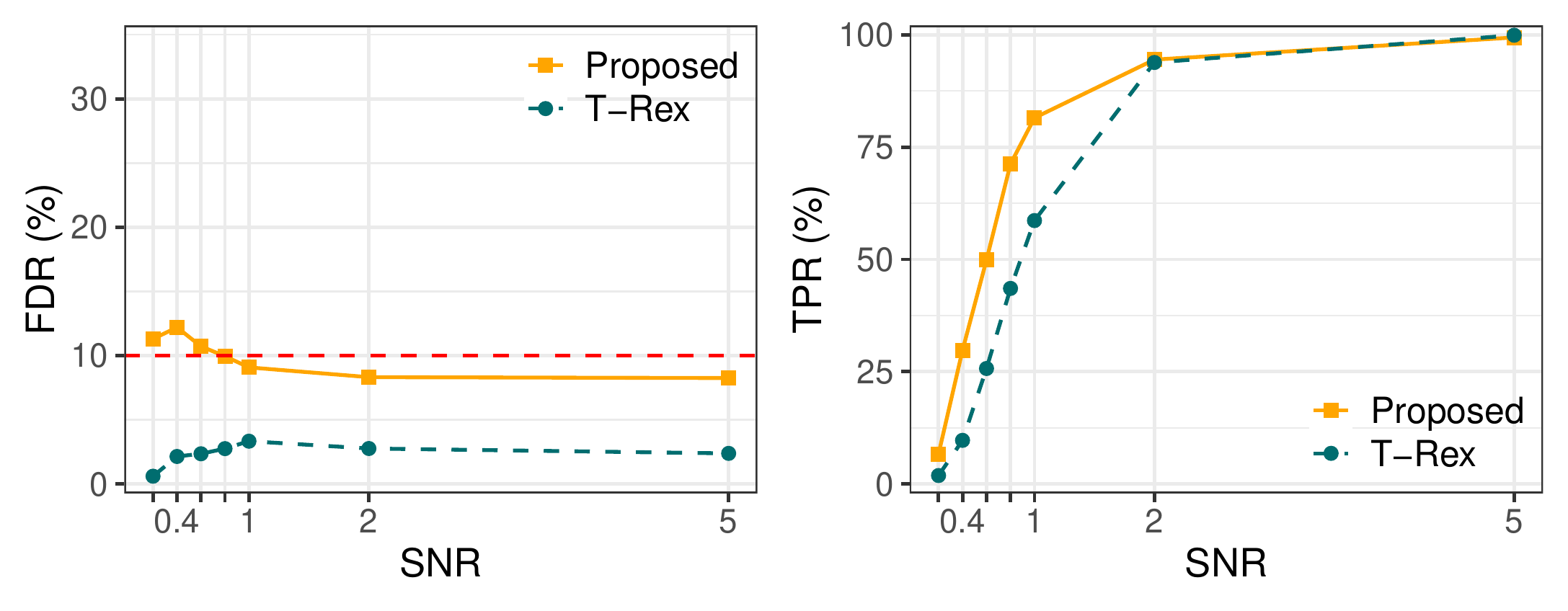}}
\caption{Comparison of original T-Rex Selector and our learning-based enhancement for different test SNR levels. Neural network is trained with multiple distributions, tested on the held-out Gaussian mixture models.}
\label{fig:fdr_tpr}
\end{figure*}

\begin{figure}[htb]
\begin{minipage}[b]{1.0\linewidth}
  \centering
  \centerline{\includegraphics[width=8.4cm]{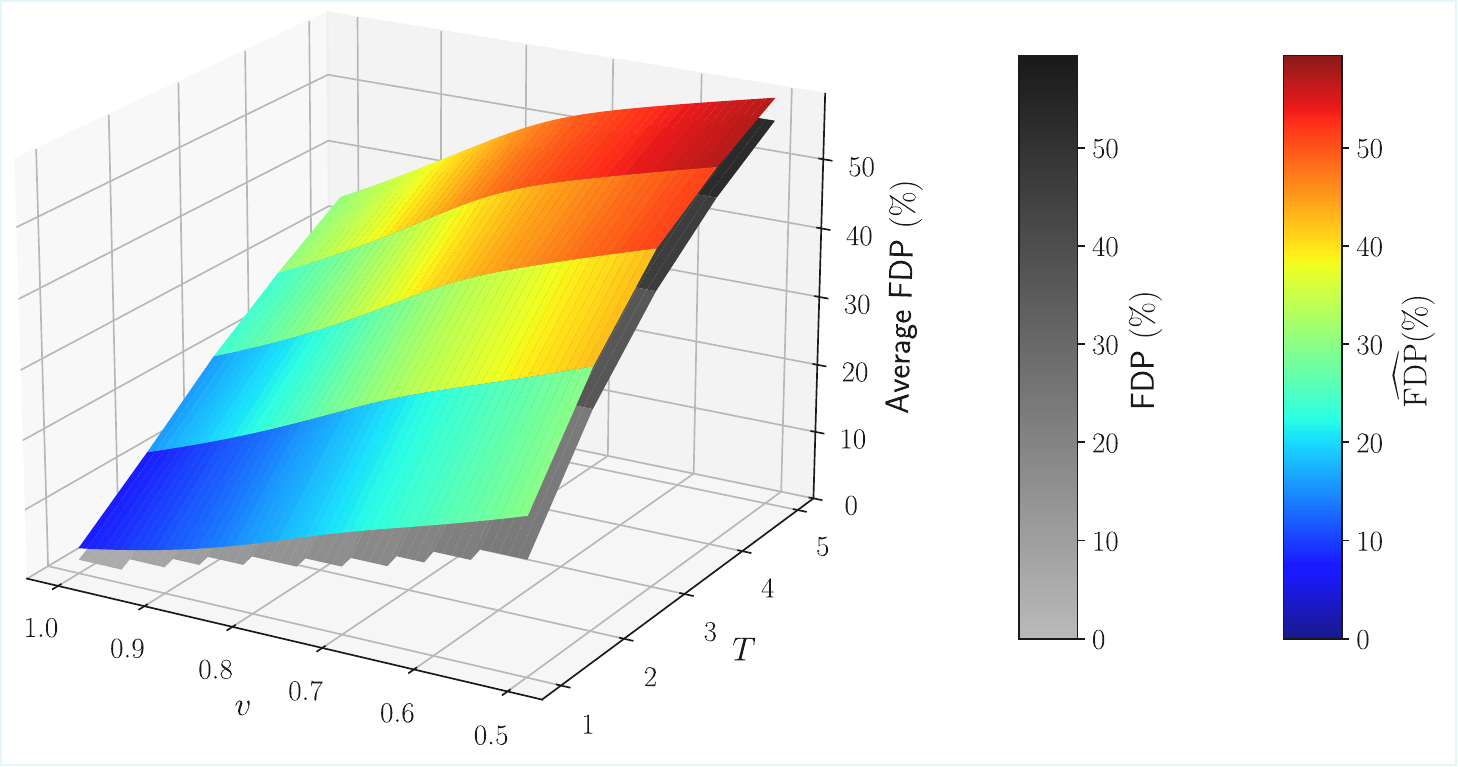}}
\end{minipage}
\caption{Average $\widehat{\text{FDP}}_{L=p}(v, T)$ produced by the neural network compared to the real $\text{FDP}$. This case achieves overestimation at all points, guaranteeing $\text{FDR}$ control.}
\label{fig:surface_plot}
\end{figure}

The empirical results demonstrate that our method successfully trades the formal FDR guarantee for a significant gain in statistical power while maintaining approximate FDR control. As shown in Fig.~\ref{fig:fdr_tpr}, our method consistently achieves a higher TPR than the original T-Rex selector across all SNR levels. This power increase is driven by the selection of less conservative calibration parameters $(v^\star, T^\star)$, particularly a lower voting threshold. This is made possible by the high fidelity of the learned FDP estimator, which provides a much tighter and generally conservative approximation of the true FDP compared to the original's analytical one (Fig.~\ref{fig:surface_plot}). Although this approximate FDR control can be less strict in the low SNR regime due to occasional underestimation, the data-driven approach is validated by the substantial power gains in most scenarios.

\vspace{-10 pt}
\section{Pushing the power on Genomics data}
\label{sec:hapgen}
\vspace{-5 pt}
To assess the practical utility and generalization of our method, we evaluate its performance on a challenging, realistic dataset from a simulated genome-wide association study (GWAS). We generate $100$ datasets using the HAPGEN2 software~\cite{su2011hapgen2}, which mimics the complex correlation structures found in real human genomic data. This provides a robust test case, as the network is evaluated on a data distribution fundamentally different from the synthetic distributions it was trained on.

Our experimental setup replicates the framework from the original T-Rex Selector paper for a direct comparison \cite{machkour2025trex}. We simulate a case-control study with $n=300$ samples ($100$ controls and $200$ cases), an initial set of $p=1,000$ predictors (SNPs), and a true active set of $|\mathcal{A}|=10$ disease-causing SNPs. The response vector $\y$ is binary, encoding the case-control status. A standard preprocessing pipeline, including SNP pruning, is applied to reduce multicollinearity (i.e., linkage disequilibrium), resulting in an average of $523$ predictors. For a comprehensive description of the data generation and preprocessing steps, we refer the reader to Appendix J in \cite{machkour2025trex}. To handle the variable number of predictors post-pruning, the network’s input dimension is fixed at the maximum observed size of $p=545$, with smaller feature sets being zero-padded.

On this genomics data, we set a target FDR of $20$\%. The results are averaged over $100$ HAPGEN2 datasets. Our method increases statistical power, boosting the TPR to $17.3\% \pm 8.2\%$ (mean $\pm$ standard deviation) from the original T-Rex selector's $14.9\% \pm 9.4\%$~\cite{machkour2025trex}. This gain is achieved while maintaining effective FDR control, with an empirical FDR of $20.8\% \pm 27.3\%$ (vs. the $15.2\% \pm 22.2\%$ obtained by the T-Rex). Our method shows approximate FDR control, only exceeding the target by a small margin.

\vspace{-5 pt}
\section{Conclusion}
\label{sec:conclusion}
\vspace{- 5pt}
We have introduced a learning-augmented enhancement of the T-Rex Selector that addresses its inherent conservativeness by replacing the analytical FDP estimator with a neural network–based approximation. This tighter estimator enables calibration closer to the target FDR, substantially increasing statistical power while preserving effective approximate control. A key advantage of our framework is its exclusive reliance on diverse synthetic training data, which ensures fairness, generalizability, and scalability without requiring labeled real-world datasets. Our experiments on both synthetic benchmarks and realistic GWAS data confirm that the proposed approach consistently boosts true discovery rates with minimal computational overhead. These results establish data-driven calibration as a powerful complement to traditional model-based methods in high-dimensional variable selection. Future directions include exploring richer neural architectures and adaptive training schemes to further enhance performance and extend applicability to a broader range of signal processing domains.

\clearpage
\bibliographystyle{IEEEbib}
\bibliography{refs}

\end{document}

%% file: packages.tex
\usepackage[T1]{fontenc}
\usepackage[utf8]{inputenc}
\usepackage[ngerman,english]{babel}
\usepackage{amsmath, amssymb, amsfonts}   
\usepackage{cuted} 
\usepackage{graphicx}         
\usepackage{algorithmic}
\usepackage{algorithm}
\usepackage{stfloats}
\usepackage{url}
\usepackage{verbatim}
\usepackage{graphicx}
\usepackage{cite}

\usepackage[nice]{nicefrac}

\usepackage{upgreek}			  
\usepackage{mathtools}		  
\usepackage{url}					
\usepackage{enumitem}		
\usepackage{accents}			
\usepackage{bbm}
\usepackage{xcolor}
\usepackage{lipsum}
\usepackage{dirtytalk}
\usepackage{xcolor}
\usepackage{float}
\usepackage{accents}
\usepackage[export]{adjustbox}
\usepackage{setspace}

	\usepackage[caption=false,font=normalsize]{subfig}

\usepackage{pifont} 
\usepackage{caption}

\usepackage{booktabs, tabularx}
\usepackage{array}

\usepackage{amsthm}
\usepackage{enumitem}

\usepackage{tikz}
\usetikzlibrary{decorations.markings,chains,matrix}
\usetikzlibrary{intersections,calc}
\usepackage{tikz-3dplot} 
\usepackage{pgfplots}
\usepackage{pgfplotstable}
\pgfplotsset{compat = newest}
\usepackage{datatool}
\usepackage{textcomp}
\usetikzlibrary{shapes,arrows}
\usetikzlibrary{decorations.pathreplacing,angles,quotes}

\usepackage[pscoord]{eso-pic} 

%% file: macros.tex
\DeclareMathOperator{\FDP}{FDP}

\newtheorem{innercustomgeneric}{\customgenericname}
\providecommand{\customgenericname}{}
\newcommand{\newcustomtheorem}[2]{%
  \newenvironment{#1}[1]
  {%
   \renewcommand\customgenericname{#2}%
   \renewcommand\theinnercustomgeneric{##1}%
   \innercustomgeneric
  }
  {\endinnercustomgeneric}
}

\newcustomtheorem{customthm}{Theorem}
\newcustomtheorem{customlemma}{Lemma}
\newcustomtheorem{customcor}{Corollary}
\newcustomtheorem{customassumption}{Assumption}

\newcommand{\y}{\boldsymbol{y}}

\newcommand{\X}{\boldsymbol{X}}

\newcommand{\D}{\boldsymbol{D}}

\newcommand{\XWK}{\boldsymbol{\widetilde{X}}}

\newcommand{\C}{\mathcal{C}}

\makeatletter
\let\save@mathaccent\mathaccent
\newcommand*\if@single[3]{%
  \setbox0\hbox{${\mathaccent"0362{#1}}^H$}%
  \setbox2\hbox{${\mathaccent"0362{\kern0pt#1}}^H$}%
  \ifdim\ht0=\ht2 #3\else #2\fi
  }
\newcommand*\rel@kern[1]{\kern#1\dimexpr\macc@kerna}
\newcommand*\widebar[1]{\@ifnextchar^{{\wide@bar{#1}{0}}}{\wide@bar{#1}{1}}}
\newcommand*\wide@bar[2]{\if@single{#1}{\wide@bar@{#1}{#2}{1}}{\wide@bar@{#1}{#2}{2}}}
\newcommand*\wide@bar@[3]{%
  \begingroup
  \def\mathaccent##1##2{%
    \let\mathaccent\save@mathaccent
    \if#32 \let\macc@nucleus\first@char \fi
    \setbox\z@\hbox{$\macc@style{\macc@nucleus}_{}$}%
    \setbox\tw@\hbox{$\macc@style{\macc@nucleus}{}_{}$}%
    \dimen@\wd\tw@
    \advance\dimen@-\wd\z@
    \divide\dimen@ 3
    \@tempdima\wd\tw@
    \advance\@tempdima-\scriptspace
    \divide\@tempdima 10
    \advance\dimen@-\@tempdima
    \ifdim\dimen@>\z@ \dimen@0pt\fi
    \rel@kern{0.6}\kern-\dimen@
    \if#31
      \overline{\rel@kern{-0.6}\kern\dimen@\macc@nucleus\rel@kern{0.4}\kern\dimen@}%
      \advance\dimen@0.4\dimexpr\macc@kerna
      \let\final@kern#2%
      \ifdim\dimen@<\z@ \let\final@kern1\fi
      \if\final@kern1 \kern-\dimen@\fi
    \else
      \overline{\rel@kern{-0.6}\kern\dimen@#1}%
    \fi
  }%
  \macc@depth\@ne
  \let\math@bgroup\@empty \let\math@egroup\macc@set@skewchar
  \mathsurround\z@ \frozen@everymath{\mathgroup\macc@group\relax}%
  \macc@set@skewchar\relax
  \let\mathaccentV\macc@nested@a
  \if#31
    \macc@nested@a\relax111{#1}%
  \else
    \def\gobble@till@marker##1\endmarker{}%
    \futurelet\first@char\gobble@till@marker#1\endmarker
    \ifcat\noexpand\first@char A\else
      \def\first@char{}%
    \fi
    \macc@nested@a\relax111{\first@char}%
  \fi
  \endgroup
}
\makeatother

\DeclareFontFamily{U}{dutchcal}{\skewchar\font=45}
\DeclareFontShape{U}{dutchcal}{m}{n}{<-> s*[1.2] dutchcal-r}{}
\DeclareFontShape{U}{dutchcal}{b}{n}{<-> s*[1.2] dutchcal-b}{}
\DeclareMathAlphabet{\mathdutchcal}{U}{dutchcal}{m}{n}
\SetMathAlphabet{\mathdutchcal}{bold}{U}{dutchcal}{b}{n}
\DeclareMathAlphabet{\mathdutchbcal}{U}{dutchcal}{b}{n}

\newlist{steps}{enumerate}{1}
\setlist[steps, 1]{label = {Step \arabic*:}, ref = {Step \arabic*}}

\newlist{alglist}{enumerate}{1}
\setlist[alglist, 1]{label = {\arabic*.}, ref = {\arabic*}}

\newcolumntype{L}[1]{>{\raggedright\arraybackslash}p{#1}}
\newcolumntype{C}[1]{>{\centering\arraybackslash}p{#1}}
\newcolumntype{R}[1]{>{\raggedleft\arraybackslash}p{#1}}

\definecolor{dark_green}{RGB}{102,166,30}
\definecolor{applegreen}{rgb}{0.55, 0.71, 0.0}

\definecolor{dark_red}{RGB}{217,95,2}
\definecolor{bittersweet}{rgb}{1.0, 0.44, 0.37}

\definecolor{dark_yellow}{RGB}{230,171,2}
\definecolor{bananayellow}{rgb}{1.0, 0.88, 0.21}

\newcommand\longvdots[1]{\raisebox{1em}{\rotatebox{-90}{\hbox to #1 {\dotfill}}}}

%% file: TRex_framework.tex
\begin{figure*}[htb]
\begin{center}
\scalebox{0.66}{
\begin{tikzpicture}[>=stealth]

  \coordinate (orig)   at (0,0);
  \coordinate (sample)   at (4,0.5);
  \coordinate (merge)   at (7,0.5);
  \coordinate (varSelect)   at (10,0.5);
  \coordinate (tFDR)   at (15.5,-1.1);
  \coordinate (fuse)   at (13,0.5);
  \coordinate (decision)   at (15.5,0.5);
  \coordinate (increment)   at (15.5,2.9);
  \coordinate (init)   at (7,-1.95);
  \coordinate (output)   at (19,0.5);
  
  \coordinate (between_scale_rank)   at (0.5,0.31);
  \coordinate (X_prime_to_tFDR_point)   at (0.5,6);
  \coordinate (X_prime_to_tFDR_point_point)   at (9,6);
  \coordinate (X_prime_to_merge_point)   at (0.5,-2.5);
  \coordinate (center_to_tFDR_point)   at (9.8,3.7);
  \coordinate (tFDR_to_fuse_point)   at (14.00,5.2);
  \coordinate (tFDR_to_sample_point)   at (4,5);
  
  \coordinate (decision_to_selection_1)   at (15.5,4.2);
  \coordinate (decision_to_selection_2)   at (10.2,4.2);

  \coordinate (Arrow_N_GenDummy)   at (4,3.06);
  \coordinate (Arrow_X_indVar)   at (7,3.06);
  \coordinate (Arrow_y_center)   at (9.5,3.7);
  \coordinate (Arrow_targetFDR_tFDR)   at (10,5.9);
  
   \coordinate (inference_Arrow)   at (15,-1.06);
   \coordinate (fuse_Arrow)   at (16.1,2.06);
  
  \coordinate (vdots1)   at (5.5,0.4);
  \coordinate (vdots2)   at (8.5,0.4);
  \coordinate (vdots3)   at (11.5,0.4);
  
  \coordinate (fuse_node)   at (15.00,0.5);
  
  \tikzstyle{decision} = [diamond, draw, 
										minimum width=2cm, minimum height=0.5cm, node distance=3cm, inner sep=0pt]

  \node[draw, minimum width=.7cm, minimum height=4cm, anchor=center , align=center] (C) at (sample) {\rotatebox{90}{\large Generate Dummies}};
  \node[draw, minimum width=.7cm, minimum height=4cm, anchor=center, align=center] (D) at (merge) {\rotatebox{90}{\large Append}};   
  \node[draw, minimum width=.7cm, minimum height=5.5cm, anchor=center, align=center] (E) at (varSelect) {\rotatebox{90}{\large Forward Variable Selection}};
  \node[draw, minimum width=.7cm, minimum height=5.5cm, anchor=center, align=center] (H) at (fuse) {\rotatebox{90}{\large Calibrate \& Fuse}};
  \node[decision, minimum width=2.7cm, minimum height=0.4cm, anchor=center, align=center] (M) at (decision) {\large $\widehat{\FDP} > \alpha$?};
  \node[draw, minimum width=2.5cm, minimum height=.7cm, anchor=center, align=center] (N) at (output) {\large Output: \\[0.3em] \large $\widehat{\mathcal{A}}_{L}(v^{\star}, T^{\star})$};
  \node[draw, minimum width=1.5cm, minimum height=.7cm, anchor=center, align=center] (O) at (increment) {\large $T \leftarrow T + 1$};
  \node[draw, minimum width=1.5cm, minimum height=.6cm, anchor=center, align=center] (P) at (init) {\large Initialize: $T = 1$};
  \node (J) at (vdots1) {\large $\vdots$};
  \node (K) at (vdots2) {\large $\vdots$};
  \node (L) at (vdots3) {\large $\vdots$};
  
  \draw[->] (Arrow_N_GenDummy) -- node[above, pos = 0.1]{$\sim\mathcal{N}(0, 1)$} ($(C.90)$); 
  \draw[->] (Arrow_X_indVar) -- node[above, pos = 0.1]{$\X$} ($(D.90)$); 
     
  \draw[->] ($(C.0) + (0,1.5)$) -- node[above]{$\D_{1}$} ($(D.0) + (-0.7,1.5)$);
  \draw[->] ($(C.0) + (0,0.75)$) -- node[above]{$\D_{2}$} ($(D.0) + (-0.7,0.75)$);
  \draw[->] ($(C.0) + (0,-1.5)$) -- node[above]{$\D_{K}$} ($(D.0) + (-0.7,-1.5)$);
     
  \draw[->] ($(D.0) + (0,1.5)$) -- node[above]{$\XWK_{1}$} ($(E.0) + (-0.7,1.5)$);
  \draw[->] ($(D.0) + (0,0.75)$) -- node[above]{$\XWK_{2}$} ($(E.0) + (-0.7,0.75)$);
  \draw[->] ($(D.0) + (0,-1.5)$) -- node[above]{$\XWK_{K}$} ($(E.0) + (-0.7,-1.5)$);
     
  \draw[->] ($(E.0) + (0,1.5)$) -- node[above]{$\C_{1, L}(T)$} ($(H.0) + (-0.7,1.5)$);
  \draw[->] ($(E.0) + (0,0.75)$) -- node[above]{$\C_{2, L}(T)$} ($(H.0) + (-0.7,0.75)$);
  \draw[->] ($(E.0) + (0,-1.5)$) -- node[above]{$\C_{K, L}(T)$} ($(H.0) +  (-0.7,-1.5)$);
     
  \path[draw,->] ($(Arrow_y_center.0)$) -- node[left,pos=0.1] {$\y$} (center_to_tFDR_point) --  ($(E.90) + (-0.2,0)$);
 
  \coordinate (between_varSelect_fuse1)   at ($(E.0) + (2.75,1.5)$);
  \coordinate (between_varSelect_fuse2)   at ($(E.0) + (2.75,0.75)$);
  \coordinate (between_varSelect_fuse3)   at ($(E.0) + (2.75,-1.5)$);
 
  \draw[->] (tFDR) -- node[below, pos=0.1]{\large $\alpha$}(M);
  \draw[->] (H) -- (M);
  \draw[->] ($(M.0)$) -- node[above, pos=0.3]{Yes}(N);
  \draw[->] ($(M.90)$) -- node[right, pos=0.3]{No}(O);
  \path[draw,->] ($(O.90)$) -- (decision_to_selection_1) -- (decision_to_selection_2) --  ($(E.90) + (0.2,0)$);
  \draw[->] ($(P.0)$) -- ($(E.180) + (0,-2.45)$);
  
\end{tikzpicture}}
\end{center}
\vspace{-5pt}
\caption{T-Rex Selector framework \cite{machkour2025trex} provides a provably conservative estimate $\widehat{\FDP}$. This may lead to a considerable gap between the actual FDP and the target FDR level $\alpha$, which results in a potentially reduced TPP.}
\label{fig:trex_scheme}
\end{figure*}

%% file: TRex_framework_enhanced.tex
\begin{figure*}[htb]
\begin{center}
\scalebox{0.665}{
\begin{tikzpicture}[>=stealth]

  \coordinate (orig)   at (0,0);
  \coordinate (sample)   at (4,0.5);
  \coordinate (merge)   at (7,0.5);
  \coordinate (varSelect)   at (10,0.5);
  \coordinate (nn_coord) at (14.5,0.5);
  \coordinate (gridsearch_coord) at (18,0.5);
  \coordinate (output_coord) at (21.5,0.5); 

  \coordinate (between_scale_rank)   at (0.5,0.31);
  \coordinate (center_to_varSelect_point)   at (9.8,3.7);
  \coordinate (nn_input_point) at (14.3,3.7);

  \coordinate (gs_input_point) at (17.8,3.7);
  \coordinate (Arrow_alpha_to_gs) at (17.5, 3.7);

  \coordinate (Arrow_N_GenDummy)   at (4,3.06);
  \coordinate (Arrow_X_indVar)   at (7,3.06);
  \coordinate (Arrow_y_center)   at (9.5,3.7);
  \coordinate (Arrow_y_to_nn) at (14, 3.7);

  \coordinate (inference_Arrow)   at (15,-1.06);

  \coordinate (vdots1)   at (5.5,0.4);
  \coordinate (vdots2)   at (8.5,0.4);

  \tikzstyle{decision} = [diamond, draw, minimum width=2cm, minimum height=0.5cm, node distance=3cm, inner sep=0pt]

  \node[draw, minimum width=.7cm, minimum height=4cm, anchor=center , align=center] (C) at (sample) {\rotatebox{90}{\large Generate Dummies}};
  \node[draw, minimum width=.7cm, minimum height=4cm, anchor=center, align=center] (D) at (merge) {\rotatebox{90}{\large Append}};   
  \node[draw, minimum width=.7cm, minimum height=5.5cm, anchor=center, align=center] (E) at (varSelect) {\rotatebox{90}{\large Forward Variable Selection}};

  \node[draw, minimum width=.7cm, minimum height=5.5cm, anchor=center, align=center] (NN) at (nn_coord) {\rotatebox{90}{\large $\text{NN}(\Phi,v, T, L)$}};
  \node[draw, minimum width=.7cm, minimum height=5.5cm, anchor=center, align=center] (GridSearch) at (gridsearch_coord) {\rotatebox{90}{\large Grid Search}};
  \node[draw, minimum width=2.5cm, minimum height=.7cm, anchor=center, align=center] (Output) at (output_coord) {\large Output: \\[0.3em] \large $\widehat{\mathcal{A}}_{L}(v^{\star}, T^{\star})$};

  \node (J) at (vdots1) {\large $\vdots$};
  \node (K) at (vdots2) {\large $\vdots$};

  \node[align=right] (I) at (12.3, 3.7) {
    $\forall v \in [0.5, \dots, 1 - \varepsilon]$ \\
    $\forall T \in [1, \dots, T_{\max}]$ \\
  };

  \draw[->] (Arrow_N_GenDummy) -- node[above, pos = 0.1]{$\sim\mathcal{N}(0, 1)$} ($(C.90)$); 

  \draw[->] ($(C.0) + (0,1.5)$) -- node[above]{$\D_{1}$} ($(D.180) + (0,1.5)$); 
  \draw[->] ($(C.0) + (0,0.75)$) -- node[above]{$\D_{2}$} ($(D.180) + (0,0.75)$); 
  \draw[->] ($(C.0) + (0,-1.5)$) -- node[above]{$\D_{K}$} ($(D.180) + (0,-1.5)$); 

  \draw[->] ($(D.0) + (0,1.5)$) -- node[above]{$\XWK_{1}$} ($(E.180) + (0,1.5)$); 
  \draw[->] ($(D.0) + (0,0.75)$) -- node[above]{$\XWK_{2}$} ($(E.180) + (0,0.75)$); 
  \draw[->] ($(D.0) + (0,-1.5)$) -- node[above]{$\XWK_{K}$} ($(E.180) + (0,-1.5)$); 

  \path[draw,->] ($(Arrow_y_center.0)$) -- node[left,pos=0.1] {$\y$} (center_to_varSelect_point) -- ($(E.90) + (-0.2,0)$);

  \path[draw,->] ($(Arrow_y_to_nn.0)$) -- node[left,pos=0.1] {} (nn_input_point) -- ($(NN.90) + (-0.2,0)$);

  \draw[->] (E) -- node[above]{$\Phi$} (NN); 


  \path[draw,->] ($(Arrow_alpha_to_gs.0)$) -- node[left,pos=0.1] {$\alpha$} (gs_input_point) -- ($(GridSearch.90) + (-0.2,0)$);
  
  \draw[->] (NN) -- node[above, midway, align=center, text width=2.5cm]{$\widehat{\text{FDP}}(\Phi,v, T, L)$} (GridSearch);
  \draw[->] (GridSearch) -- node[above]{$v^{\star}, T^{\star}$} node[below]{\small $L$} (Output);

  \def\xLeftBox{10.5}
  \def\xRightBox{23.25} 
  \def\yTopBox{4.5}
  \def\yBottomBox{-2.5}

  \draw[dashed, -] (\xLeftBox, \yTopBox) -- (\xLeftBox, \yBottomBox);
  \draw[dashed, -] (\xLeftBox, \yBottomBox) -- (\xRightBox, \yBottomBox);
  \draw[dashed, -] (\xRightBox, \yBottomBox) -- (\xRightBox, \yTopBox);
  \draw[dashed, -] (\xRightBox, \yTopBox) -- (\xLeftBox, \yTopBox);

  \node[anchor=north east] at (\xRightBox, \yBottomBox-0.05) {\small Proposed};

\end{tikzpicture}}
\end{center}
\vspace{-20pt}
\caption{T-Rex Selector framework with the proposed enhancement using a neural network.}
\label{fig:proposed_scheme}
\end{figure*}